\ifdefined\SIAMTypeSetting
\documentclass{siamltex1213}
\else
\documentclass[notitlepage,twocolumn,superscriptaddress,floatfix]{revtex4-1}
\fi

\usepackage{graphicx,color}  %
\usepackage{amssymb,amsmath}
\usepackage{amsfonts}

\graphicspath{{./}}

\begin{document}

\ifdefined\SIAMTypeSetting
\title{FAST AND ACCURATE EVALUATION OF WIGNER {3-j}, {6-j}, and {9-j}
  SYMBOLS USING PRIME FACTORISATION AND MULTI-WORD INTEGER ARITHMETIC}
\author{H.~T.~Johansson\footnotemark[2]\ \footnotemark[5]
\and C.~Forss\'en\footnotemark[2]\ \footnotemark[3]\ \footnotemark[4]\ \footnotemark[5]}

\maketitle  

\else
\title{Fast and accurate evaluation of Wigner 3j, 6j, and 9j symbols using
  prime factorisation and multi-word integer arithmetic}
\author{H.~T.~Johansson}
\affiliation{Department of Fundamental Physics, Chalmers University of
  Technology, SE-412 96 G{\"o}teborg, Sweden} 
\author{C.~Forss\'en} 
\affiliation{Department of Fundamental Physics, Chalmers University of
  Technology, SE-412 96 G{\"o}teborg, Sweden} 
\affiliation{Department of Physics and Astronomy, University of
  Tennessee, Knoxville, TN 37996, USA}
\affiliation{Physics Division, Oak Ridge National Laboratory, Oak Ridge,
  TN 37831, USA} 

\fi

\begin{abstract}
  We present an efficient implementation for the evaluation of Wigner
  $3j$, $6j$, and $9j$ symbols. These represent numerical
  transformation coefficients that are used in the quantum theory of
  angular momentum. They can be expressed as sums and square roots of
  ratios of integers. The integers can be very large due to
  factorials. We avoid numerical precision loss due to cancellation
  through the use of multi-word integer arithmetic for exact
  accumulation of all sums. A fixed relative accuracy is maintained as
  the limited number of floating-point operations in the final step
  only incur rounding errors in the least significant bits. Time spent
  to evaluate large multi-word integers is in turn reduced by using
  explicit prime factorisation of the ingoing factorials, thereby
  improving execution speed.  Comparison with existing routines shows
  the efficiency of our approach and we therefore provide a computer
  code based on this work.
\end{abstract}

\ifdefined\SIAMTypeSetting
\renewcommand{\thefootnote}{\fnsymbol{footnote}}
\footnotetext[2]{Department of Fundamental Physics, Chalmers University of
  Technology, SE-412 96 G{\"o}teborg, Sweden}
\footnotetext[3]{Department of Physics and Astronomy, University of
  Tennessee, Knoxville, TN 37996, USA}
\footnotetext[4]{Physics Division, Oak Ridge National Laboratory, Oak Ridge,
  TN 37831, USA}
\footnotetext[5]{The research leading to these results has received funding from the
European Research Council under the European Community's Seventh
Framework Programme (FP7/2007-2013) / ERC grant agreement no.~240603,
and the Swedish Foundation for International Cooperation in Research
and Higher Education (STINT, IG2012-5158). This material is
based upon work supported in parts by the U.S.\ Department of Energy
under Contract No.\ DE-AC05-00OR22725 (Oak Ridge National
Laboratory). The authors would like to
acknowledge valuable comments on the manuscript and the computer code
by B.\ D.\ Carlsson and K.\ Wendt.}

\renewcommand{\thefootnote}{\arabic{footnote}}

\pagestyle{myheadings}
\thispagestyle{plain}
\markboth{H.~T.~JOHANSSON AND C.~FORSS\'EN}{FAST AND ACCURATE EVALUATION OF WIGNER n-j SYMBOLS}

\else

\maketitle  

\fi

\section{Introduction}
\label{sec:intro}

Wigner $3j$, $6j$, and $9j$ symbols ($xj$ symbols) are used in physics and
chemistry whenever one deals with angular momenta in quantum
mechanical systems~\cite{varshalovich1988quantum,brink1993angular}.
These functions correspond to transformation coefficients between
different spin representations. The input arguments are six or nine
integers, or half integers, that represent different angular momenta $j$
and projection quantum numbers $|m| \le j$.
Because of their frequent appearance in scientific computing, standard
library functions that evaluate Wigner symbols are available in
several mathematics software systems~\cite{sympy,mathematica,sage} and
computational libraries~\cite{Galassi:2013wz}.
In fact, the quantum spin properties of electrons, protons, neutrons
and atoms imply that angular momentum recoupling is a general and
important ingredient in the theoretical modeling of quantum
many-body systems~\cite{fetter2003quantum}.  Well known examples are
found in quantum chemistry~\cite{atkins2011molecular}, condensed
matter~\cite{dickhoff2008many}, atomic~\cite{lindgren2012atomic}, and
nuclear physics~\cite{suhonen2007nucleons}.
More recently, the mathematical apparatus of quantum-mechanical
angular momentum coupling has found a prominent role also in quantum
gravity and quantum computing applications~\cite{Aquilanti2008}.

In particular, the Wigner symbols enter at a critical stage in the computational chain of
quantum many-body modeling; namely when calculating the strength of the
interaction between pairs or triples of particles. The evaluation of
these so called interaction matrix elements can involve intermediate
angular-momentum sums that extend to very large values ($j \sim
100$). Eventually, a sum over all possible particle pairs or triples
within the many-body system gives the total interaction energy.
The continued efforts to improve both accurate
calculations~\cite{Stone:1980bt,Wei:1999hc,Anderson2008}, constructing
recursive relations~\cite{schulten1975,Luscombe:1998dr,Tuzun:1998kk},
and achieving fast look-up~\cite{Rasch:2003dr} show the central place
of these symbols in computations.

Accurate evaluation of Wigner symbols with large $j$ using
floating-point arithmetic is difficult because of cancellation in
sums of large alternating terms.  The precision losses due to
cancellations can be avoided by using integer arithmetic. However,
this approach generally involves very large numbers due to factorials,
thus requiring the use of multi-word integer representations in
computer codes.  Multi-word integers are multiple machine words
used together to represent arbitrarily large numbers.
A clever reduction of the huge integers, by
expressing the terms as products of binomial coefficients, was
demonstrated by L.~Wei~\cite{Wei:1999hc}. In this context, one can
also mention early efforts introducing prime
factorisation~\cite{Dodds1972,Stone:1980bt}.
However, high-accuracy approaches are in general significantly slower
than straight-forward floating-point implementations. As an example, the
computation of $6j$ symbols for small angular momenta, $\max(j) = 3$,
is already an order of magnitude slower when using the published code
of Ref.~\cite{Wei:1999hc} as compared to a floating-point calculation
with the popular GNU Scientific Library
(GSL)~\cite{Galassi:2013wz}. This difference is growing with
increasing $j$.

In this paper we present a new method in which the big-integer
penalties are mitigated by making explicit use of prime-factorised
factorials. In our tests this method executes considerably faster than
previous high-accuracy approaches, provides a bounded and fixed
relative accuracy, and extends to very large values of ingoing angular
momenta.

\section{Method}
The implementation uses the fact that the $xj$ symbols can be written
on the form
\begin{equation}
  \label{exprnsd}
  W_x = \frac{n \sqrt{s}}{q},
\end{equation}
where $n$, $s$ and $q$ are integers.  This general expression comes
directly from the defining equations~\cite{varshalovich1988quantum}.
E.g., for $6j$ symbols we have the Racah formula
\begin{multline}
  \label{expr6j}
  \left \{
    \begin{array}{ccc}
      a & b & c \\
      d & e & f \\
    \end{array}
  \right \}
  =
  \prod_{i=1}^4 \sqrt{\Delta(\hat{\alpha}_i)}
\\
\times
  \sum_k (-1)^{k} 
\frac{(k+1)!}
  {\prod_{i=1}^4 (k-\alpha_i)!
    \prod_{j=1}^3 (\beta_j-k)!},
\end{multline}
with
\begin{equation}
\Delta(a,b,c) = \frac{(a+b-c)! (a-b+c)! (b+c-a)!}{(a+b+c+1)!},
\end{equation}
and 
\begin{equation}
  \begin{array}{l@{\quad}l@{\quad}l}
    \hat{\alpha}_1 = (a,b,c), & \alpha_1 = a+b+c, & \beta_1 = a + b + d + e, \\
    \hat{\alpha}_2 = (d,e,c), & \alpha_2 = d+e+c, & \beta_2 = a + c + d + f, \\
    \hat{\alpha}_3 = (a,e,f), & \alpha_3 = a+e+f, & \beta_3 = b + c + e + f, \\
    \hat{\alpha}_4 = (d,b,f), & \alpha_4 = d+b+f.&
  \end{array}
\end{equation}
The summation index $k$ is in the range $\max(\alpha_1, \alpha_2,
\alpha_3, \alpha_4) \le k \le \min(\beta_1, \beta_2, \beta_3)$.  As
stated before, the input arguments $a,
b, c, d, e,$ and $f$ are integers or half-integers.  For a symbol
not to be trivially zero, the sums $\alpha_i$ must be integer and
$\hat{\alpha}_i$ must fulfill triangle conditions, e.g.\ for
$\hat{\alpha}_1$:
\begin{equation}
  |a - b| \le c \le a + b.
\end{equation}
The overall structure for $3j$ symbols is similar.  Note how all
expressions contain (factorials of) integers.  While being
possibly large, they can be evaluated exactly.  By finding and using
the least common denominator (LCD) of the sum, expression
\eqref{expr6j} can be brought to the form \eqref{exprnsd}.

Finding the LCD is straightforward by expressing each factorial as an
explicit product of prime numbers, e.g.\ $7! = 2^4 \times 3^2 \times
5^1 \times 7^1$, treated as the tuple $(4,2,1,1)$ in the calculations.
The tuple representing each term is found by subtracting the exponents
related to each prime number in all factors of the denominator from
the ones representing the numerator.  The LCD is then found by taking
the minimum value for each exponent over all terms. In practice, the
LCD is reduced considerably since the  numerator is also included in
this procedure. Factors common to all terms are then represented by
positive exponents.
As an explicit example we consider the computation of the $6j$ symbol
$\left\{ \begin{smallmatrix} 2 & 2 & 2 \\ 2 & 2 & 2 \end{smallmatrix}
\right\}$.
The sum that needs to be evaluated for this symbol includes three
terms, see Table~\ref{tab:6jexample}.  The final common factor of the sum 
(which includes the LCD) is
found to have the tuple representation $(1,2,1,1)$. The positive
exponents are a consequence of the relatively large numerators
appearing in the sum.

The common factor is then subtracted from the exponent tuples of each term,
giving pure numerators.  In Table~\ref{tab:6jexample} this step
corresponds to the second to last column. They are converted into
exact integers using multi-word arithmetics.  The sum is then easily
accumulated, yielding one part of $n$ in Eq.~\eqref{exprnsd}.
\begin{table*}
  \caption{\label{tab:6jexample}Evaluation of the $6j$ symbol $\left\{
      2 \, 2 \, 2 ; 2 \, 2 \, 2  \right\}$
    that illustrates the identification of the common factor (including
    the LCD) and the accumulation
    of the numerator sum. The fifth column corresponds to the tuple
    representation of each term, i.e.\ the numerator 
    tuple subtracted by the denominator product tuple. See text for details.}
\begin{center}
\ifdefined\SIAMTypeSetting
\footnotesize
\renewcommand{\arraystretch}{1.3}
\fi
\begin{tabular}{lcccc|cr}
\hline
Term & Sign & Numerator & Denominator & Total ratio &
  \multicolumn{2}{c}{Final numerator} \\
& & & & (prime tuple) & (prime tuple) &
(integer) \\
\hline 
$k=6$ & $+1$ & $7!$ & $2!2!2!$ & $(1,2,1,1)$ & $(0,0,0,0)$ & $1$ \\
$k=7$ & $-1$ & $8!$ & $1$ & $(7,2,1,1)$ & $(6,0,0,0)$ & $-64$ \\
$k=8$ & $+1$ & $9!$ & $2!2!2!2!$ & $(3,4,1,1)$ & $(2,2,0,0)$ & $36$ \\
\hline
\hline
\multicolumn{4}{r}{Common factor / LCD:} & $(1,2,1,1)$ &
  \multicolumn{1}{|r}{Sum:} & $-27$ \\
\hline
\end{tabular}
\end{center}
\end{table*}

The product of the $\Delta$-factors is formed by summing the factorial
prime exponent contributions.  As it shall suffer a square root, one
factor will be removed from each odd exponent to make sure that all
remaining ones are even. The removed odd ones will build the term
$\sqrt{s}$ in Eq.~\eqref{exprnsd}, while the remaining even ones will be
divided by two, thereby evaluating the square root. In our example,
the product under the square root becomes
$\frac{2!^3}{7!}  \frac{2!^3}{7!} \frac{2!^3}{7!} \frac{2!^3}{7!}$
that is represented by the tuple
${(-1,-2,-1,-1)^4} = {(-4,-8,-4,-4)}$. All exponents are even so
applying the square root simply gives $(-2,-4,-2,-2) \sqrt{1}$.

The common factor is then added, giving final exponents that are used to
create $q$ and the second part of $n$ in Eq.~\eqref{exprnsd}, by using
the negative and positive exponents, respectively. In our example, the
product of the $\Delta$-factors is added to the common factor of
Table~\ref{tab:6jexample} giving the final exponents $(-1,-2,-1,-1)$,
which corresponds to $1/630$.  The two parts of $n$ are combined using
multi-word integer multiplication.

In the final step, the value of $W_x$ in (\ref{exprnsd}) can be
evaluated using floating-point arithmetics with a limited loss in
relative precision.  In our example, this corresponds to the
evaluation of $-27 \sqrt{1} / 630 = -0.0429$. The three conversions
from integer to floating point, together with one square root, one
multiplication, and one division constitute six operations that each
incur the loss of (at most) half a least significant bit in the
floating-point representation used, commonly referred to as the
machine epsilon, $\epsilon$.  For the 64-bit floating-point format of
IEEE 758, which is usually used for the \texttt{C} type
\texttt{double}, one has $\epsilon = 1.11\cdot 10^{-16}$. This implies
a maximum relative error of $6\epsilon = 6.66\cdot 10^{-16}$. 

\subsection {Wigner 9j%
\label{wig9j}}
Wigner $9j$ symbols can be written as a sum of products of $6j$
symbols, see e.g.\ Ref.~\cite{varshalovich1988quantum}. It is not
obvious from this direct formula for Wigner $9j$ symbols that they can
be written on the form \eqref{exprnsd} due to the square roots in the
$\Delta$ prefactors of the $6j$ symbols.
Nevertheless, a rearranged formula based on binomials was
presented~\cite{Wei:1998hf} and used~\cite{Wei:1999hc} by L.~Wei.
This formula is also employed in the present implementation, with the
small modification of expressing the binomials as factorials.  In
total, the $9j$ evaluation becomes a double sum where an overall LCD
expressed on prime-exponent form can be extracted. The routine
implementing the sum in the $6j$ formula is re-used for this purpose.

\section{Results%
\label{sec:results}}
\begin{figure*}
  \centering
    \includegraphics[width=0.9\textwidth]{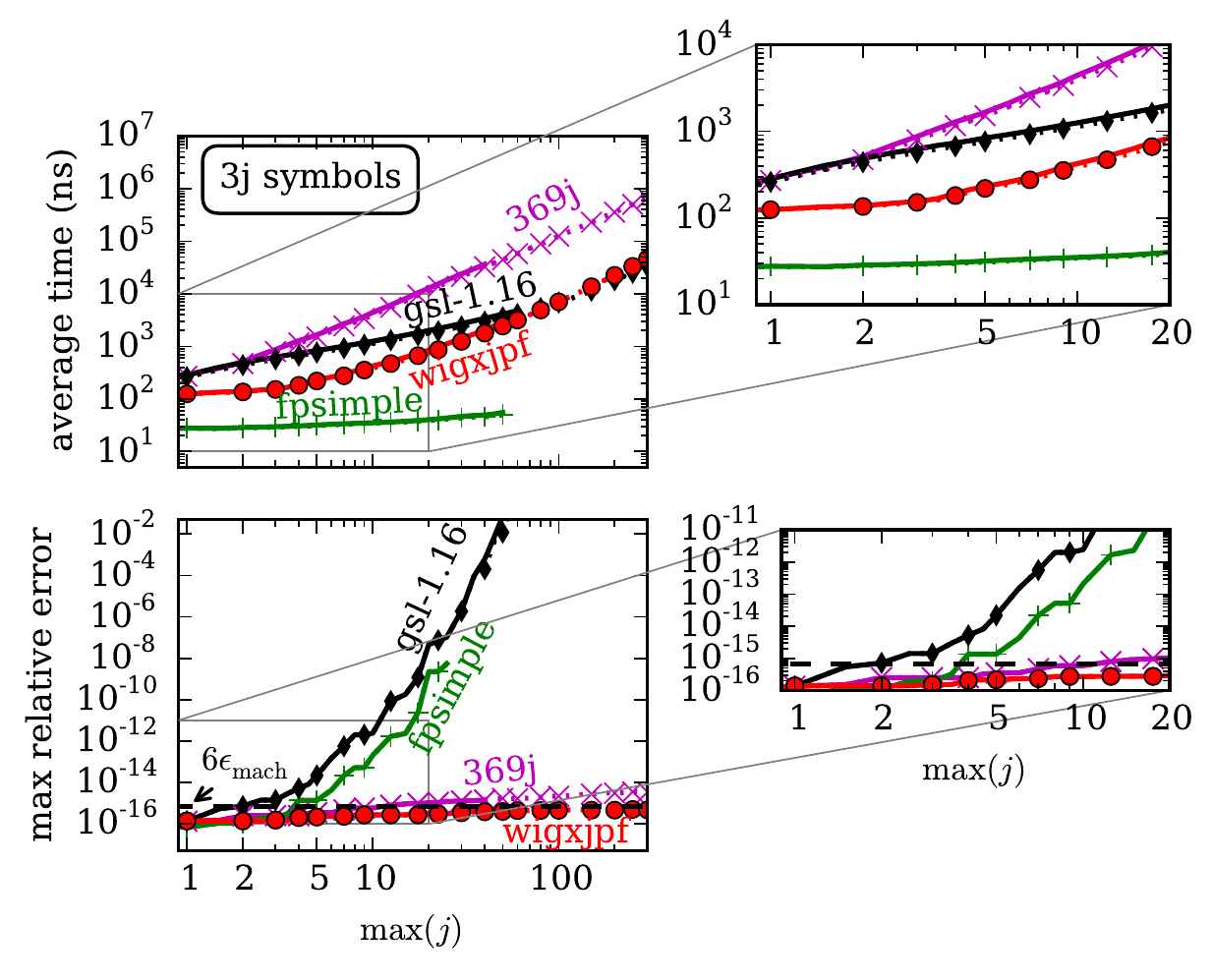}
    \caption{\label{fig:3j} Upper panel: average evaluation time for
      $3j$ symbols using four different implementations:
      \texttt{wigxjpf} (this work, circle markers),
      \texttt{369j}~\cite{Wei:1999hc} (crosses),
      GSL-1.16~\cite{Galassi:2013wz} (diamonds), and a simple
      floating-point computation (plus markers).  The solid lines
      correspond to the exhaustive measurements where all
      non-trivially-zero symbols with a
      specific maximum $j$ have been computed. The dotted lines
      represent an approximation to the average evaluation time
      obtained from computations of a random subset.  The overhead
      time to enumerate symbols, etc., has been subtracted. Lower
      panel: maximum relative error, as compared to the symbols
      evaluated using \texttt{long double} on x86 hardware.  All tests
      have been performed using a Xeon E3-1240v3, single-threaded @
      3.8 GHz. The dashed line indicates the expected maximum relative
      error of our implementation, which is six times the machine
      epsilon.}
\end{figure*}
Calculations using the \texttt{wigxjpf} routine presented in this work
have been compared (in floating point) with the \texttt{369j} code by
Wei~\cite{Wei:1999hc} and with a simple implementation in which we use
floating-point operations from precalculated factorial tables (see
Sec.~\ref{sec:implementation}).  The floating-point based routines of
GSL (version 1.16)~\cite{Galassi:2013wz} have also been
included in the measurements.  The comparisons are both in terms of
execution speed and accuracy.
\begin{figure*}
  \centering
    \includegraphics[width=0.9\textwidth]{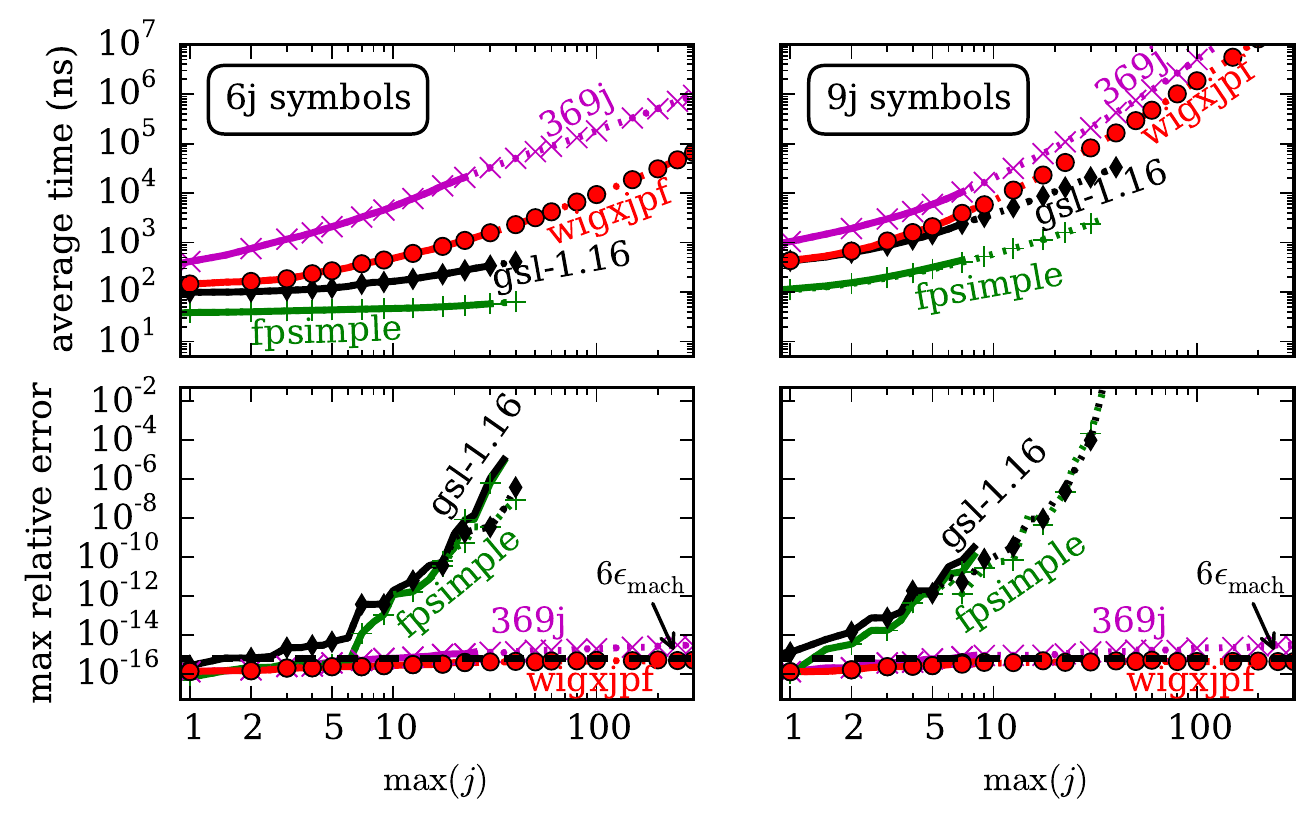}
    \caption{\label{fig:6j9j} Same as Fig.~\ref{fig:3j} but for the
      evaluation of $6j$ (left panels) and $9j$ symbols (right
      panels).}
\end{figure*}
The benchmark calculations for measuring the accuracy are performed
with our routine, \texttt{wigxjpf}, using extended precision
computations (the 80 bit \texttt{long double} of the x87
floating-point unit, $\epsilon = 5.42\cdot 10^{-20}$). This benchmark
is reliable since only the last floating-point stage of the current
routine introduces rounding errors.

Results are presented in Fig.~\ref{fig:3j} and Fig.~\ref{fig:6j9j} for
$3j$ and $(6j,9j)$ symbols, respectively.  For reasonable maximum $j$,
all symbols not trivially zero by triangle conditions have been
calculated (solid lines), which provides an exhaustive measurement of the
average evaluation speed and the maximum relative error.  However, for
larger symbols the measurement is approximative as we have restricted
the computations to a random subset of non-trivially-zero symbols
(dotted lines with markers).

In all measurements, the present work is slightly more accurate and up
to an order of magnitude faster than \texttt{369j}.  The relative
error never exceeded $6 \epsilon \approx 6.66\cdot 10^{-16}$, which is
our upper bound.  The slightly larger accuracy loss in \texttt{369j}
is due to individual conversion of each prime power to floating point
and subsequent multiplication.
The execution time difference between our implementation and the fast
floating-point based routines is less than an order of magnitude at
low $j$, showing that the \texttt{wigxjpf} routines presented in this
work are also very efficient. In comparison to GSL (version 1.16),
\texttt{wigxjpf} is actually faster for $3j$ symbols up to
$\max(j)=60$, while the difference in execution speed always remains
smaller than a factor 4 up to $\max(j)=20$ (30) for $6j$ ($9j$) symbols.
At large $j$ the floating-point routines obviously suffer from an
increasing loss of accuracy, thus making \texttt{wigxjpf} superior.

Explicit numerical values, execution times, and memory usage for a
selected set of symbols are presented in Table~\ref{tab:values}. Note
that we present results for extreme cases such as $j=50,000$ and
$j=2,000$ for $6j$ and $9j$ symbols, respectively.
\begin{table*}
  \caption{\label{tab:values}Values and execution times for selected $xj$
    symbols.  These times include setup of the precalculated
    tables of factorised factorials.  The memory used for those tables and
    other temporary storage is also given.  All tests have been
    performed using a Xeon E3-1240v3, single-threaded @ 3.8 GHz.}
\begin{center}
\ifdefined\SIAMTypeSetting
\footnotesize
\renewcommand{\arraystretch}{1.3}
\fi
\begin{tabular}{lcr@{.}lr@{.}lr@{ }l}
\hline
 \multicolumn{2}{c}{Symbol}  & \multicolumn{2}{c}{Value} &
 \multicolumn{2}{c}{Time}    & \multicolumn{2}{c}{Memory} \\
\hline
3j & $\left( \begin{array}{ccc}
 15  &  30   & 40   \\
 2   &  2   & -4   \\
 \end{array} \right)$        & -0&01908157979919155  & 0&07~ms  & 58&kB   \\
   & $\left( \begin{array}{ccc}
 200 &  200 & 200  \\
 -10 &  60  & -50  \\
 \end{array} \right)$        & 0&0007493927313989515 & 0&31~ms  & 730&kB \\
   & $\left( \begin{array}{rrr}
 50{,}000 &  50{,}000 & 50{,}000  \\
  1{,}000 &  -6{,}000 &  5{,}000  \\
 \end{array} \right)$        & -1&116843916927519e-05 & \multicolumn{2}{l}{112~s} & 19&GB \\
\hline
6j & $\left\{ \begin{array}{ccc}
 8   &  8   & 8   \\
 8   &  8   & 8   \\
 \end{array} \right\}$       & -0&01265208072315355  & 0&06~ms  & 7.0&kB  \\ 
   & \{ all 200 \}           & 0&0001559032124132416 & 0&62~ms  & 1.0&MB  \\
   & \{ all 600 \}           & -1&03981778344144e-07 & 6&1~ms   & 8.0&MB  \\
   & \{ all 10,000 \}         & 2&770313640470537e-08 & 8&2~s & 1.5&GB  \\
   & \{ all 50,000 \}         & 3&997351841910046e-08 & \multicolumn{2}{l}{816~s} & 32&GB  \\
\hline
9j & $\left\{ \begin{array}{ccc}
 8.5 &  9.5 & 7.0 \\
12.5 &  8.0 & 8.5 \\
 8.0 & 10.5 & 9.5 \\
 \end{array} \right\}$       & 0&0002812983019125448 & 0&08~ms & 20&kB \\
   & $\left\{ \begin{array}{ccc}
100 & 80 & 50 \\
50 & 100 & 70 \\
60 & 50 & 100 \\
 \end{array} \right\}$       & 1&055977980657612e-07 & 1&9~ms   & 0.50&MB \\
   & \{ all 200 \}           & 1&278335300545066e-07 & 0&15~s   & 1.6&MB  \\
   & \{ all 1,000 \}          & 1&749851385596156e-09 & 29&8~s  & 30&MB \\
   & \{ all 2,000 \}          & 2&755181565857189e-10 & \multicolumn{2}{l}{358~s} & 109&MB \\
\hline
\end{tabular}
\end{center}
\end{table*}

\section{Implementation%
\label{sec:implementation}}
Before the calculation of individual symbols, a table with
precalculated factorisations of factorials is prepared.  Table
\ref{tab:max} shows the maximum possible factorial that has to be
computed for evaluation of different symbols.  The time spent on this
initialization stage is amortized for applications that make repeated
use of the routines.  The table also gives the maximum number of terms
in the sum (inner sum for $9j$ symbols), for each of which temporary
results (numerator-denominator exponent tuples) are stored during
execution.
\begin{table}
  \caption{\label{tab:max}Maximum factorial argument and 
    number of terms given the maximum $j$ in a symbol.  The limits can
    be rounded down when non-integer.  These limits must be considered
    for temporary memory allocation. The memory requirement for the
    precalculated factorial tables are given in the last columns for
    three different values of $\max(j)$.}
\begin{center}
\ifdefined\SIAMTypeSetting
\footnotesize
\renewcommand{\arraystretch}{1.3}
\fi
\begin{tabular}{c|cc|ccc}
\hline
& \multicolumn{2}{|c|}{Maximum} & \multicolumn{3}{c}{Memory requirement}  \\
Symbol & $p!$ & terms &  
$j=100$ & $j=1{,}000$ & $j=10{,}000$ \\
\hline 
3j & $(3j+1)!$ & $j+1$ & 193 kB & 10.8 MB & 0.78 GB \\
6j & $(4j+1)!$ & $j+1$ & 309 kB & 17.9 MB & 1.35 GB\\
9j & $(5j+1)!$ & $j+1$ & 450 kB & 27.5 MB & 2.06 GB\\
\hline
\end{tabular}
\end{center}
\end{table}

First, all prime numbers up to the largest factorial argument, $p$,
are determined by the sieve of Eratosthenes.  By simple enumeration 
of integer exponent combinations in
the basis of prime numbers, the factorisation of all integers up to
$p$ are determined.
Thus the factorisations are
performed without any 'test' divisions, i.e., checks for remainder zero.
Each factorisation is stored as an array of
integers.  The factorisation of factorials are constructed by
cumulating the previous factorisations.

Multiplying (or dividing) two such factorisations is a
matter of adding (or subtracting) each element of the arrays, giving a
new array.
Adding or subtracting two factorisations would be more complicated as the
result, expressed as a factorisation, could be completely different,
even requiring much larger prime factors.  To avoid this, each term is
converted from the prime-factor array representation to multi-word
integer before addition or subtraction.  No attempt is made to extract
common factors between different multi-word integers, as this is
expensive and would have no impact on accuracy.

The multi-word integers are just multiple machine words (32 or 64
bits) used to represent arbitrarily large numbers.  The highest word
is treated as having a sign bit, while all others are unsigned.
Routines that perform addition,
subtraction and multiplication are implemented.
The main difficulty is to propagate overflow (i.e.\ carry) between
the words.
The execution times
for addition and subtraction are linear in the number of words used,
while multiplication is
quadratic.  

The conversion of a number in prime-factor form to multi-word integer
form evaluates the contribution for each prime number separately,
before multiplying them together.  We utilize the fact that, for each
exponent, each successively higher bit in its binary number
representation corresponds to the square of the value represented by
the previous bit.  Therefore, two values are kept while iterating
through the bits of each exponent: the repeatedly squared prime number
and a cumulative product.  While non-zero bits remain, the first
number is squared each iteration,
starting as the prime number for the first bit.
For each bit set to one, the cumulative product is multiplied by the
hitherto squared result.  These intermediate results are treated as
multi-word integers when necessary.

Conversion from multi-word integer to floating point in the final stages
is done word by
word, from lower to higher words.  The contributions that cannot be
represented (i.e.\ that are below the precision) in the resulting
floating-point variable are therefore just lost/ignored.

Repeated similar calculations of factorial arguments
can be avoided by noting that the factors (and
denominators) of each successive term in the sum depend on the
summation index $k$ in such a way that they always change their
argument by 1 for each iteration.  Therefore, factorial arguments
(table addresses) can be calculated before the sum loop. In the actual
loop these addresses are then just moved up or down one step each
iteration.

Symbols that are trivially zero can be identified using a single
\texttt{if}-clause in combination with two's-complement arithmetics
and bit manipulation.
Only symbols passing this test are evaluated.

For the comparisons presented in Sec.~\ref{sec:results} we have also
implemented simple floating-point routines. They use precalculated
floating-point factorial values employing the same list-address
technique for the sum as described above and thus become very fast.  
The factorials, stored as
floating-point values, are calculated precisely using the multi-word
integers.  The accuracy loss is completely dominated by cancellation
in the sums of large alternating terms.

\subsection {Memory usage%
  \label{sec:implementation:memory}} 
The largest memory requirement of \texttt{wigxjpf} is associated
with the tables of precalculated factorisations of numbers and
factorials, see Table \ref{tab:max}.
The next largest use is the temporary arrays of prime
factors of each term in the sum. Other variables that are used during
computations, e.g.\ multi-word integers, are comparatively small in
size.

\section{Conclusions and Outlook%
\label{concl}}
In conclusion, we have successfully implemented a computational method
for fast and accurate evaluation of Wigner $3j$, $6j$, and $9j$
symbols using prime factorisation and multi-word integer arithmetics. 
The efficiency and accuracy of our approach has been validated via
benchmark calculations and comparison with existing floating-point
routines and with the \texttt{369j} code by Wei~\cite{Wei:1999hc}. The
latter algorithm is similar to our approach in the sense that it also
uses prime-number factorisation and multi-word integer
arithmetics. However, it employs recursive calculation of binomial
coefficients while our implementation, with the precalculation of a
table with prime-number factorisations, offers a code that is
significantly faster; comparable in speed even to the floating-point
routines. In addition, the precomputed factorisation table makes our
approach particularly amenable to applications in which multiple
evaluations of angular momentum coupling coefficients are needed.

Furthermore, our implementation produces results with a maximum
relative error of only six times the machine epsilon for $3j$, $6j$, and
$9j$ symbols, also for very large angular momenta. In
comparison, the error in the evaluation of $6j$ symbols using GSL 1.16
is five orders of magnitude larger already for $j=15$ and quickly grows
with increasing values of the angular momenta.

A computer code with the routines presented in this work,
\texttt{wigxjpf}, is available for download~\cite{chalmers:wigxjpf}.

\ifdefined\SIAMTypeSetting
\bibliographystyle{siam}
\else
\section*{Acknowledgments}

The research leading to these results has received funding from the
European Research Council under the European Community's Seventh
Framework Programme (FP7/2007-2013) / ERC grant agreement no.~240603,
and the Swedish Foundation for International Cooperation in Research
and Higher Education (STINT, IG2012-5158). This material is
based upon work supported in parts by the U.S.\ Department of Energy
under Contract No.\ DE-AC05-00OR22725 (Oak Ridge National
Laboratory). The authors would like to
acknowledge valuable comments on the manuscript and the computer code
by B.\ D.\ Carlsson and K.\ Wendt.
\fi

\bibliography{wigxjpfpaper,wigxjpfpaper_temp}

\end{document}